\RequirePackage{fix-cm}
\documentclass[smallextended]{svjour3}       % onecolumn (second format)
\smartqed  % flush right qed marks, e.g. at end of proof
\usepackage{natbib}
\usepackage{graphicx}
\begin{document}

\title{Dynamical capture in the Pluto-Charon system%\thanks{Grants or other notes
}
%\subtitle{ progenitors
 % of the small satellites}


\author{P. M. Pires dos Santos        \and
        A. Morbidelli \and
        D. Nesvorn{\'y} %etc.
}


\institute{P. M. Pires dos Santos \at
              UNESP-Universidade Estadual Paulista, Campus de Guaratinguet{\'a}, Brasil \\
  %            Tel.: +55-12-31232800\\
   %           Fax: +55-12-31232800\\
              \email{pos09032@feg.unesp.br}           %  \\
           \and
           A. Morbidelli \at
              Departement Lagrange: Universit\'e de Nice - Sophia Antipolis, Observatoire de la C\^{o}te d'Azur, CNRS, France\\
           \email{morby@oca.eu} 
           \and
          D. Nesvorn{\'y} \at
              Department of Space Studies, Southwest Research Institute, 1050 Walnut St, Suite 400, Boulder, CO 80302, USA\\
           \email{davidn@boulder.swri.edu}
}

\date{Accepted: 24 August 2012}

\maketitle

\begin{abstract}
This paper explores the possibility that the progenitors of the small
satellites of Pluto got captured in the Pluto-Charon system from the
massive heliocentric planetesimal disk in which Pluto was originally
embedded into.  We find that, if the dynamical excitation of the disk
is small, temporary capture in the Pluto-Charon system can occur with
non-negligible probability, due to the dynamical perturbations exerted
by the binary nature of the Pluto-Charon pair. However, the captured
objects remain on very elliptic orbits and the typical capture time is
only $\sim 100$~years. In order to explain the origin of the small satellites of
Pluto, we conjecture that some of these objects got disrupted during
their Pluto-bound phase by a collision with a planetesimal of the
disk. This could have generated a debris disk, which damped under
internal collisional evolution, until turning itself into an
accretional disk that could form small satellites on circular orbits,
co-planar with Charon.  Unfortunately, we find that objects large
enough to carry a sufficient amount of mass to generate the small
satellites of Pluto have collisional lifetimes orders of magnitude
longer than the capture time. Thus, this scenario cannot explain the
origin of the small satellites of Pluto, which remains elusive.


\end{abstract}

\section{Introduction}
\label{intro}
The Pluto-Charon-Nix-Hydra system is in the Kuiper Belt, a disk of
numerous objects spanning the region beyond Neptune's orbit up to
$\sim$50~AU.  Charon, the largest satellite, has a radius about half
of Pluto's radius, which implies that the center of mass of the
Pluto-Charon system is outside Pluto.  The outer satellites Nix and
Hydra, discovered in 2005 \citep{weaveretal2006}, are much smaller than
Charon and they orbit the center of mass on nearly circular orbits, in
the orbital plane of Charon.

The four-body orbital solution of \citet{tholenetal2008} determined 
accurate masses of Charon, Nix and Hydra.  The best-fit Charon/Pluto
mass ratio is 0.1166, which implies that Charon's mass is 1.52 $\times
10^{21}$~kg.  The estimated masses of Nix and Hydra are 5.8$\times
10^{17}$~kg and 3.2$\times 10^{17}$~kg, respectively.  The diameters
of both small satellites were estimated by assuming that they have a
Charon-like density (1.63~g~cm$^{-3}$). This gives diameters of 88~km
and 72~km for Nix and Hydra, respectively.

The location of Nix suggests that it is near the 4:1 mean motion
resonance with Charon, while Hydra is near the 6:1 resonance
\citep {weaveretal2006, buieetal2006}. However, the investigation of
several different resonant arguments showed that Nix and Hydra are not
in 4:1 and 6:1 resonances with Charon, respectively, at present
\citep{tholenetal2008}.  Any model that aims at explaining the origin of
Nix and Hydra should explain how these bodies ended up in their
near-resonant orbits with small eccentricities and small inclinations
related to Charon's orbital plane.

The formation of Pluto-Charon pair through a giant impact \citep{mckinnon1989, canup2005} is widely accepted.
In the past few years, the origin of Nix and Hydra has been debated in
the literature. 
\citet{sternetal2006}, \citet{wardcanup2006} and
\citet{canup2011} advocated a scenario in which the origin of Charon, Nix
and Hydra is credited to the same event, i.e. as the result of a large collision of an ancient body on Pluto. 
The collision left Charon, the remnant of the projectile, on an orbit with a semi
major axis equal to a few Pluto radii \citep{canup2005}, and generated a
debris disk just beyond Charon's orbit \citep{canup2011}. The disk was not
radially extended enough to account for the currently large orbital
radii of the satellites Nix (43 radius of Pluto, $R_p$) and Hydra
(57~$R_p$) \citep[see Figure 6][]{canup2011}. However, Charon migrated by
tidal interaction to its actual position at $\sim$17~$R_p$. In this
process, it was proposed that Nix and Hydra got captured in the 4:1
and 6:1 mean motion resonances with Charon and then they were transported
outwards as these resonances migrated together with Charon. 

It remains to be shown, however,
whether resonant migration can efficiently transport Nix and Hydra.
Migration in a mean motion
resonance typically raises the eccentricities of the resonant
particle, while the current orbits of Nix and Hydra are basically
circular. To overcome this problem, \citet{wardcanup2006} proposed
that Nix and Hydra got captured in the 4:1 and 6:1 co-rotation
resonances with an eccentric Charon. In fact, co-rotation resonances do
not excite the particles eccentricities during the outward migration.
However, the set of parameters that allows capture in the 4:1
co-rotation resonance and the one that allows capture in the 6:1
resonance have an empty intersection: so, the \citeauthor{wardcanup2006}
mechanism could not have worked for Nix and Hydra simultaneously
\citep{lithwickwu2008}.  Also, the transport of Nix and Hydra's
orbits by resonant migration is ruled out by extensive numerical
simulations of tidal models of Charon's migration \citep[e.g.][]{pealeetal2011,cheng2011}; most of the resonant particles are
not stable and are eventually scattered by Charon and removed. When
the hydrostatic value $J_2$ of Pluto is taken into account, disk
particles can not be transported in resonance while preserving near
circular orbits and all the test particles are eventually ejected.

In the last year another small satellite was discovered orbiting Pluto
between the orbits of Nix and Hydra. This ``new'' body, temporarily
named P4, has a diameter estimated between 13 to 34~km. This satellite
was discovered during a search for faint dust rings; however, no such ring was discovered$\footnote[1]{For the information
  regarding the discovery of P4, visit:
  http://hubblesite.org/newscenter/archive/releases/2011/23/}$. During the preparation of this manuscript
a fifth moon  was found orbiting Pluto$\footnote[2]{The discovery of the latest moon is reported in
http://hubblesite.org/newscenter/archive/releases/solar\%20system/2012/32/full/}$, hereafter P5.

In this paper, we consider a potential alternative scenario for the
origin of the minor satellites of Pluto and their peculiar orbits.  We
briefly explain here the general idea.  

In the early phases of the Solar System, the Pluto-Charon binary was
embedded in a massive disk of planetesimals, probably 1,000 times more
populated than the current Kuiper Belt \citep[see][for a review]{morbyetal2008}.
This massive disk might have survived for hundreds of
millions of years, before being strongly depleted by the orbital
evolution of the giant planets \citep{levisonetal2008}. During the
massive disk phase, numerous planetesimals should have had close
encounters with the Pluto-Charon binary. If Pluto had been a single
object, all encounters would have been hyperbolic fly-bys.  However, the
existence of Charon opens the possibility of three-body (even
four-body, including the effect of the Sun) energy exchanges, leading
to the capture of the incoming planetesimal on a bound orbit around
Pluto. The captured orbits are expected to have large eccentricities
and all captures are expected to be temporary, because energy
exchanges are reversible. However, if the captured planetesimal(s) had been
broken by collisions with other incoming objects during their
capture-phase, then the disk of debris generated by the break-up
could have behaved in a dissipative way, damping the debris eccentricities
and inclinations by mutual collisions. Eventually a debris disk
co-planar with Charon and with circular orbits could have formed, leading
subsequently to the formation of small satellites such as Nix, Hydra
and P4,5.

The plausibility of this idea needs to be investigated on quantitative
grounds. Thus, in section 2, we compute the probability that
planetesimals encountered the Pluto-Charon binary during the
massive-disk phase as well as the orientations and mutual velocities
of their incoming orbits.  We then explain how we use this information
to set up capture simulations.  In section 3 we discuss the results of
these capture simulations and estimate the size of the largest
planetesimals that should have experienced a temporary capture around
Pluto. We then compute the collisional lifetimes of the captured
planetesimals as a function of size and compare them with their
dynamical lifetimes as Pluto-bound objects. Section 4 summarizes our
conclusions.

\section{Method}
\label{sec:1}

As we said in the introduction, we consider an early phase during
which Pluto was embedded in a massive planetesimal disk. Because Pluto
was one of the most massive objects in the disk, presumably its orbit
had a small eccentricity and a small inclination relative to the
mid-plane of the disk, as a consequence of dynamical friction
\citep{wetherillstewart1993}. So, we can assume for simplicity that
both Pluto's eccentricity and inclination were null. It is not known
where Pluto was before the events that depleted the primordial
trans-Neptunian disk and formed the current Kuiper Belt (placing
Pluto onto its current orbit). There is a consensus that it was much
closer to the Sun than it is today \citep{malhotra1993, levisonetal2008}.  Without loss of generality, we assume that the orbit of
Pluto had a semi-major axis of 20 AU.

The first step of our investigation is to compute the encounter
probability, velocity and orientation of the disk's planetesimals
relative to the Pluto-Charon binary. For this purpose, we assume that
the eccentricities $e$ of the disk's planetesimals were randomly
distributed up to $e_{\rm max}$ (a free parameter, whose value is
discussed in sect. 3), while the inclinations $i$ spanned the interval
(0--$i_{\rm max}$), with $i_{\rm max}=e_{\rm max}/2$. We also assume
that the semi major axes $a$ of the planetesimals were randomly
distributed in the interval 18 to 22 AU. We generate in this way sets
of $a,e,i$ for the disk's planetesimals, until we find a number $M$
$\sim$ 1200 planetesimals crossing the assumed orbit of the
Pluto-Charon binary.

\begin{figure}[!htb]
\begin{center}
 \includegraphics[scale=.45]{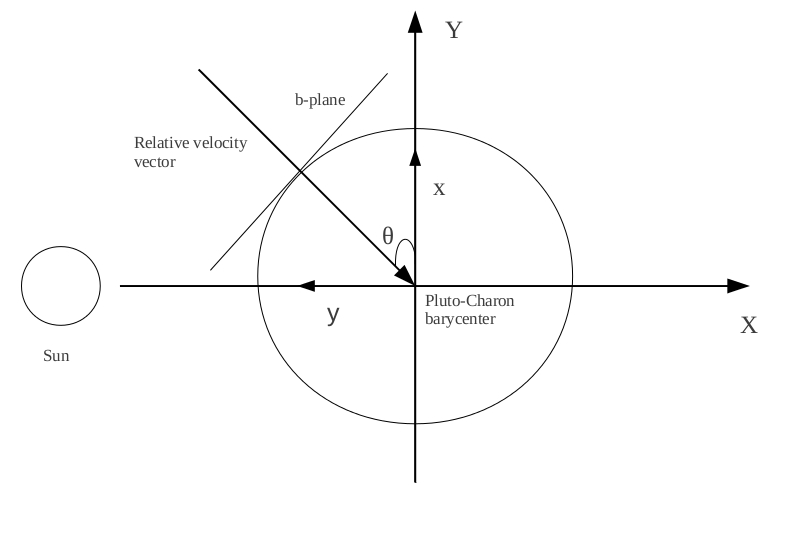}\\
\hspace{0.50cm} \includegraphics[scale=.5]{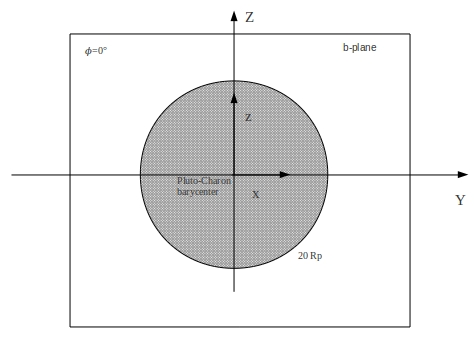}
\caption{From the top to bottom: view of the reference plane x-y and a representation of the b-plane when $\phi= 0^{\circ}$. The line connecting
the Sun, Pluto and Charon defines the X-axis.}
\label{angs}
\end{center}
\end{figure}

For each of these planetesimals, we compute the intrinsic collision
probability $p$ (defined as the 
probability per target km$^2$, per year), the unperturbed relative velocity $v$ and the
orientation of the relative velocity vector ($\theta, \phi$). This is
done using a Opik-like approach \citep{wetherill1967}. The orientation
angles $(\theta,\phi)$ are defined in a reference frame where the
$x$-axis is directed along the velocity vector of the Pluto-Charon
binary relative to the Sun, while the $y$-axis is directed towards the
Sun from the Pluto-Charon barycenter (Fig.~\ref{angs}). More
precisely, $\theta$ is the angle defining the projection of the
relative velocity vector on the $x$--$y$ plane and $\phi$ gives the
latitude of the relative velocity vector relative to that plane.  In
fact, there are 4 possible equi-probable encounter configurations,
corresponding to the 4 possible combinations of the signs of $\theta$
and $\phi$.  For each planetesimal, we chose randomly these two signs,
thus fixing the encounter geometry.

For each planetesimal, we then simulate the close encounter of the
Pluto-Charon binary with a swarm of $N$ particles, with $N$
proportional to the value of $p$ for the considered
planetesimal. We use $N=1$ for the planetesimal with the smallest
$p=9.3 \times 10^{-21}$km$^{-2}$~yr$^{-1}$. The particles are uniformly distributed
on the $b$-plane \citep{valsecchimanara1997}, all having the same
velocity vector relative to the Pluto-Charon binary as the considered
planetesimal.  The $b$-plane is here defined as the plane orthogonal to
the relative velocity vector, which is tangent to the Hill sphere of
the Pluto-Charon binary (Fig.~\ref{angs}).  On the $b$-plane we set a
new coordinate frame. The center of the frame is the projection of
the Pluto-Charon barycenter along the direction of the relative
velocity vector.

Because the gravitational focusing of the Pluto-Charon system is small
for the relative planetesimal velocities considered in this work (see
sect. 3), the trajectories of the incoming particles are weakly
curved. Thus, particles passing more than 20 Pluto radii away from
the center of the frame on the $b$-plane have no chance to be
deflected by encounters with Charon (which is at $\sim$15$R_p$ from
the Pluto-Charon barycenter).  For this reason, we distribute the
particles uniformly on the $b$-plane, but only up to 20$R_p$ from the
reference frame center.

For the particles that constitute the swarm associated to each
planetesimal, the recipe described above sets the initial positions
and velocities relative to the barycenter of the Pluto-Charon binary
and the Sun.  To start the integrations, we now need to fix the orbit
and the position of Charon. We do this as follows. We assume that the
inclination of the orbit of Charon around Pluto, measured relative to
the orbital plane of the binary around the Sun, was the same as the
current orbital inclination of Charon relative to the ecliptic plane
119$^{\circ}$ \citep{tholenbuie1997}. This is justified because, if the
inclination of Pluto's orbit was impulsionally excited during the
dispersion of the primordial trans-Neptunian disk and the formation of
the current Kuiper Belt, the orientation of the orbital plane of
Charon should have been preserved relative to an inertial frame. We
then assume that the line of nodes of Charon orbit had a random
orientation on the ($x,y$) plane. In addition, we assume Charon to
have a random position along its circular orbit.

Finally, we apply a series of rotations and translations in order to
transform the full system (Pluto, Charon, Sun and particles of the
swarm) to a new reference frame, centered on Pluto, whose reference
plane is the orbital plane of Charon.  The integrations are performed
in this new Pluto-centric frame, using the swift\_rmvs3 integrator from
the Swift package \citep{levisonduncan1994}.  Each integration is
continued until all swarm particles exceed a distance of two Hill
radii (or 12,000~$R_p$) from Pluto (2~$R_H$ hereafter). We call ``temporarily trapped" the particles which have, at some time, a
negative energy relative to the Pluto-Charon barycenter.

The probability $P_i$ that a given planetesimal ($\#i$) is captured
onto an orbit temporarily bound to the Pluto-Charon system is then
computed as follows. Denote $K_i$ the number of particles temporarily
trapped out of the $N_i$ particles integrated in its swarm, and denote
by $p_i$ the planetesimal's intrinsic collision probability.
Then one has:
\begin{equation}
P_i= p_i(20R_p)^2 K_i/N_i \ 
\label{temp-capt1}
\end{equation}
the probability $P_i$ is expressed in yr$^{-1}$.

The mean probability of temporary trapping for the population of
Pluto-crossing planetesimals in the disk is therefore
\begin{equation}
 P=\sum_{i=1}^{M} (P_i / M).
\label{temp-capt2}
\end{equation}
where $M$ is the number of planetesimals that we studied on
Pluto-Charon-crossing orbits.

\section{Results}

\subsection{Capture event}

We started by assuming that the excitation of the disk can be
characterized by $e_{\rm max}=0.1$.  This value comes from simulations
of the self-excitation of the disk in the case where there are $\sim
1,000$ Pluto-size objects \citep{levisonetal2009, levisonetal2011}.  In this case
we find that the mean encounter velocity with the Pluto-Charon system,
weighted over the intrinsic encounter probability $p$ is
0.4~km~s$^{-1}$.

\begin{figure}[!htb]
\begin{minipage}[b]{0.5\linewidth}
\includegraphics[width=\linewidth]{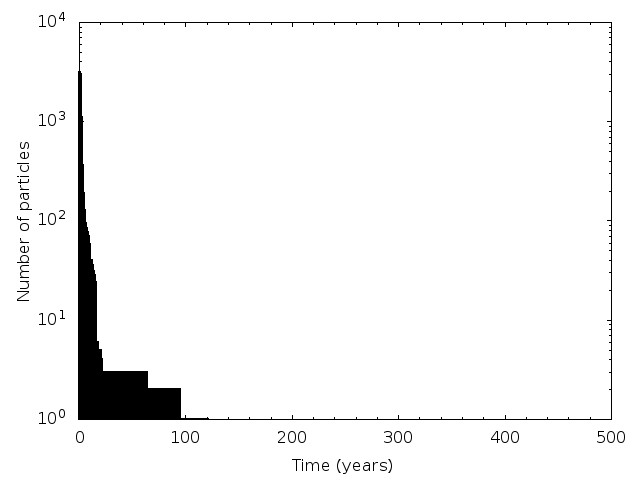}
a)
\end{minipage} \hfill
\begin{minipage}[b]{0.5\linewidth}
\includegraphics[width=\linewidth]{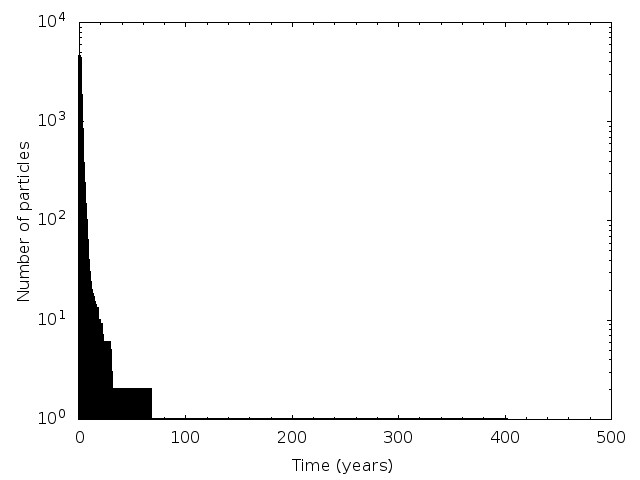}
b)
\end{minipage}
\begin{minipage}[b]{0.5\linewidth}
\includegraphics[width=\linewidth]{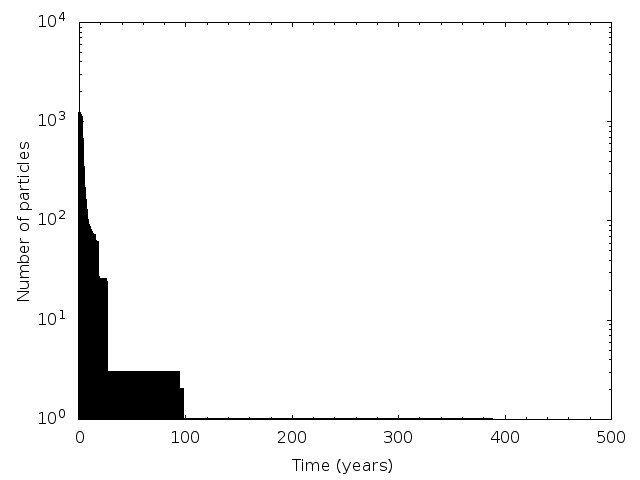}
c)
\end{minipage} \hfill
\begin{minipage}[b]{0.5\linewidth}
\includegraphics[width=\linewidth]{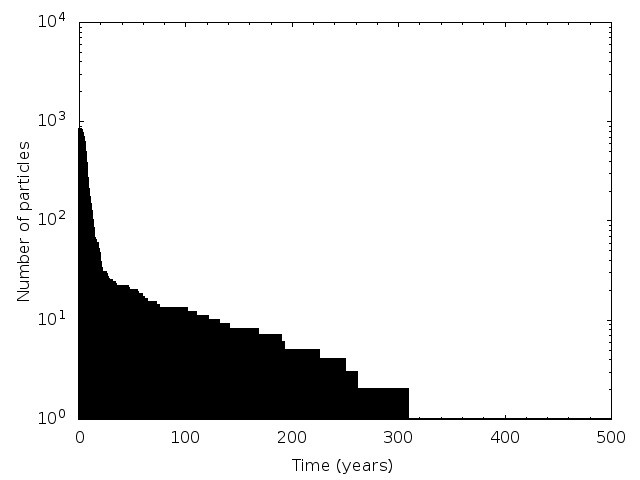}
d)
\end{minipage}
\caption{The cumulative number of particles having a lifetime longer
  than a given value, for different assumed excitations of the
  planetesimal disk: $e_{max}=$ 0.1 ( a) ), 0.07 ( b) ), 0.05 ( c) )
  and 0.03 ( d) ).}
\label{lifetime}
\end{figure}

With this characteristic incoming speed, the cumulative distribution
of particle lifetimes is that shown in Fig.~\ref{lifetime}~a. Remember
that the lifetime is measured here from the initial condition of a
particle on the $b$-plane (approximately 1 $R_H$ away from the
Pluto-Charon barycenter) to the moment when the particle's distance
from said barycenter exceeds 2$R_H$. Notice that the cumulative
distribution is made of two distinct features. There is initially
sharp decay, due to the fact that most particles have lifetimes
shorter than 20~years, followed by a ``foot'', due to a few particles with longer
lifetimes, up to $\sim 100$y. The particles with short lifetimes just
have hyperbolic fly-bys with the Pluto-Charon system. The spread in
lifetimes up to $\sim 20$y is due to the spread in incoming velocities
of the considered planetesimals.  Instead, the ``foot'' is due to
particles that experience temporary capture. In summary, more than a
half of incoming particles have a lifetime shorter than 10 years. Only
1 particle remains bound to Pluto until 120 years, however this
particle is very far from the planet, $q$ (barycentric) reaches $600
R_p$ during its orbital evolution.  This is not very promising for our
scenario.

Therefore, we considered disks with smaller dynamical excitation
($e_{\rm max}=0.07$ and 0.05).  These disks correspond to weighted
mean encounter velocities with the Pluto-Charon system of
0.3~km~s$^{-1}$ and 0.2~km~s$^{-1}$. The resulting cumulative
distributions of particle lifetimes (Fig.~\ref{lifetime}~b and c) do
not change much relative to the previous case. The initial decay is a
bit slower, because the incoming velocities are smaller. Again,
particles with lifetimes longer than $\sim 20$y experience temporary
capture, and they are very few.

Finally, we decreased the eccentricity excitation of the disk to
$e_{\rm max}=0.03$.  The distribution of cumulative lifetimes
(Fig.~\ref{lifetime}~d) changes qualitatively relative to the previous
cases.  Now, many more particles experience temporary trapping (all
those with lifetime longer than 22y), so that the ``foot'' of the
distribution is well developed and we can appreciate the distribution
of lifetimes within the ``foot''. The drastic increase in number of
temporary captured particles is due to the fact that, with a weighted
mean velocity at infinity of 0.1km/s, several particles have now a
velocity lower than Pluto's velocity around the Pluto-Charon
barycenter: 0.025km/s.  Particles passing through the Pluto-Charon
system typically experience a velocity change of this order and
therefore, if their velocity at infinity is smaller, they are likely
to be captured. 

\begin{figure}[!htb]
\begin{center} 
%\begin{minipage}[b]{0.5\linewidth}
\includegraphics[width=0.61\linewidth]{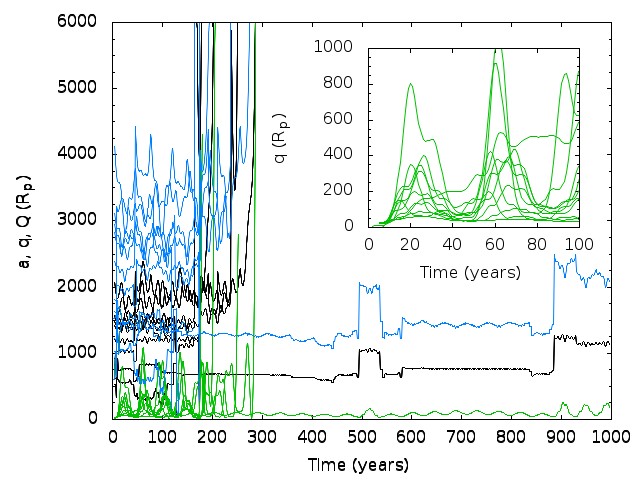}\\
%\end{minipage} \hfill
%\begin{minipage}[b]{0.5\linewidth}
\includegraphics[width=0.6\linewidth]{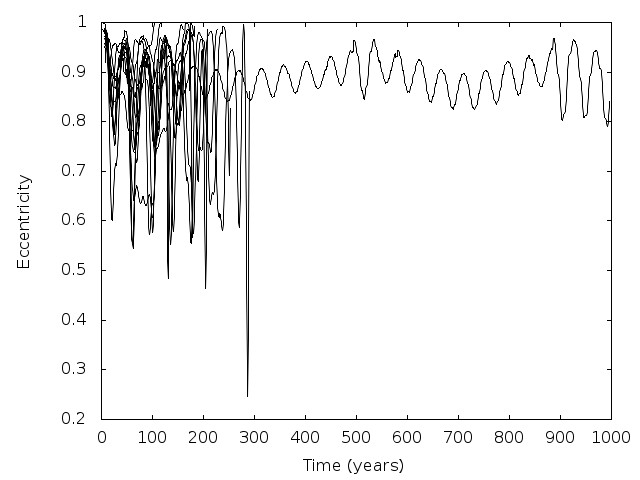}\\
%\end{minipage}
\includegraphics[width=0.6\linewidth]{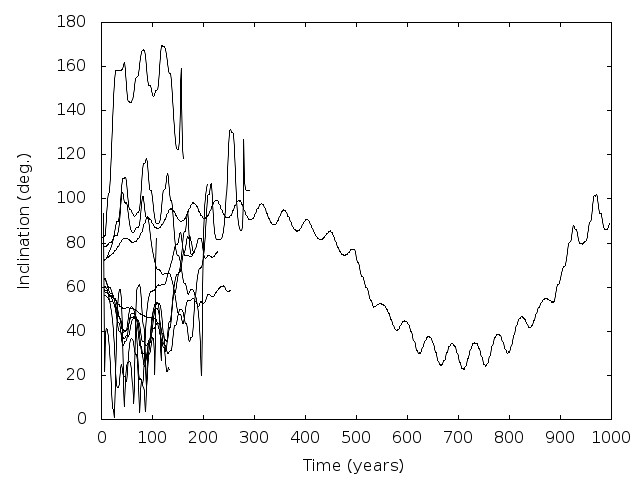}
\end{center}
\caption{Orbital evolution of particles captured by Pluto-Charon for at least 100~years. Top panel:
blue lines represent the maximum plutocentric distances (Q), black lines represent the semimajor axis (a), and green
lines represent the minimum plutocentric distances (q).
Middle and bottom panel: evolution of the eccentricity and inclination, respectively. }
\label{evol2}
\end{figure}

From this last sample, the orbital evolution, during the time range when the barycentric orbital elements are elliptic, of the
long-lived particles (bound to Pluto for at least 100~years) are shown in the Fig. \ref{evol2}.
The captured orbits cover a wide range of semimajor axis, eccentricity and inclination. Nevertheless,
the pericentre varies from 30 to $\sim$ 1000~$R_p$. Eventually, collisions nearby Pluto could happen.
In the next few years ($>$100~years)
the orbital elements change significantly, and the particles escape in hyperbolic trajectories.
As an isolated case, one particle remains trapped into elliptical orbit during the whole simulation (1000~years).

We now compute the temporary trapping probability expressed as a
fraction of the Pluto-crossing population per year.  This is done
using (\ref{temp-capt2}). We find that $P$=1.1 $\times 10^{-13}
y^{-1}$. Assuming that the massive planetesimal disk lasted about
500~My as in the Nice model \citep{gomesetal2005}, then the probability
that a Pluto-crossing particle experiences a temporary capture during
the lifetime of the disk is 5.5$\times 10^{-5}$.

Using this information and a model for the size distribution of
particles in the disk, we can now estimate the size of the largest
planetesimal that should have experienced a capture event. Obviously,
this is the diameter ($D$) for which the cumulative number of
particles $N(>D)=1/5.5\times 10^{-5}$.  As a disk model, we adopted
that defined in \citet{morbyetal2009} (see Fig. 1a of that paper),
which is consistent with all the constraints of the Nice model.  The
disk considered in \citeauthor{morbyetal2009}, though, was spread in about 15 AU
in radius. If we assume that $e_{\rm max}$ of the disk was 0.03, as to
have a significant number of captures, one gets a mean eccentricity of $e_{\rm
  max}/2=0.015$ and only approximately 4$\%$ of such a disk would cross
the orbit of Pluto. Therefore, the size of the largest temporarily
captured planetesimal is the one such that $N(>D)/25=1/5.5\times
10^{-5}$. We find $D=300$km.

A body of this size carries a mass that is 25 times that of Nix and
Hydra combined. Therefore, if this body were broken during its
temporary capture phase, enough material would be liberated as a disk
of debris around Pluto to allow, potentially, the formation of the
small satellites observed today. Therefore, in the next section we
compute the collisional lifetime of planetesimals in our adopted disk
and compare it with the capture time.

Notice that, in principle, one should consider also the possibility
that planetesimals are collisionally disrupted during hyperbolic
fly-bys because, even if most fragments would escape from the
Pluto-Charon system in that case, a fraction of them could be captured
thanks to the wide distribution of the fragment ejection
velocities. However, we checked that the cumulative time spent in
hyperbolic fly-bys in all distributions of Fig.~\ref{lifetime} is less
than the cumulative lifetime of temporary captured particles in the
distribution of Fig.~\ref{lifetime}~d. Thus, investigating the
probability of disruption of temporary captured particles, as we do in
the next section, is enough to test our model. In fact, if this
probability turns out to be too small, the probability of collisional
disruption during a hyperbolic fly-by would be even smaller.

\subsection{Collisional disruption estimate}

For a planetesimal with radius $R=150$km, we assume that the value of
the specific catastrophic disruption energy $Q^*$ is the one given by
\citet{benzasphaug1999} for competent ice and
$v_{imp}=0.5$~km~s$^{-1}$. The mean impact velocity in our adopted
disk is smaller than 0.5~km~s$^{-1}$.  Moreover, it is possible that
primordial trans-neptunian objects were weaker than competent ice
planetesimals by a factor of 4 \citep{leinhardtstewart2009}.
However, we are interested here in an order of magnitude estimate of
the collisional lifetime to see whether our proposed scenario is
plausible or not. Thus, we believe that the use of Benz $\&$ Asphaug
is enough for our purposes at this stage.

In general, assuming equal bulk densities, the size of a projectile
required to catastrophically disrupt a target is:

\begin{equation}
D_p= (2Q^*/v^2_{\rm imp})^{1/3} D_t,
\end{equation}
where $D_p, D_t$ are the diameters of projectile and target,
respectively, and $v_{\rm imp}$ is the impact speed.

In our case, $Q^* > 3 \times 10^8$~ergs~g$^{-1}$ and $v_{\rm imp}$ =
10$^4$ cm~s$^{-1}$ (because captured particles are very eccentric --see
Fig.~\ref{evol2}-- and spend most of the time near aphelion we neglect
here the gravitational focusing of the Pluto-Charon system and the
orbital motion of the target). Thus $(2Q^*/v^2_{\rm imp})>1$. This
means that the target cannot be broken by anything smaller than its
own size. It can only be broken if it smashes into something bigger,
but this is out of the validity regime of the Benz and Asphaug formula
(in fact, in this case the body would be a projectile, not a
target). This means that, in a dynamically cold disk such as the one
that we have been forced to assume to observe some temporary captures,
collisions of $D=D_t$ bodies are accretional, for $D_t=300$km.

The maximal size of bodies that can be collisionally broken in a
collision with an equal-size body is the one such that $Q^*(D)=(1/2)
v^2_{\rm imp}$.  Using again Fig. 7 from \citet{benzasphaug1999}, we
find $D=20$~km. According to Fig.~1 from \citep{morbyetal2009}
these should be approximately 10$^{10}$ objects of this size or larger
in the disk, of which 2.2$\times 10^4$ would have experienced
temporary capture in the Pluto-Charon system (as we said above, 4\% of
the particles would cross Pluto's orbit and $5.5\times10^{-5}$ would be
captured). Notice that this ensemble of bodies with $D$~=~20~km
contains $\sim$160 times the cumulative mass of Nix and Hydra,
assuming equal densities.

We now estimate the collisional lifetime of $D$=20~km bodies. Using the Opik-like approach, we compute that 
the collision intrinsic probability, averaged over all crossing particles is $\bar{p}= 2.04\times10^{-19}$~km$^{-2}$~y$^{-1}$. The probability of
collisional break-up per year is therefore

\begin{equation}
 P^*= \bar{p} [(D_p+D_t)/2]^2 N(>D_p)
\end{equation}
where $N(>D_p)$ is the number of objects larger than $D_p$ crossing
the orbit of the target.  For $D_t=D_p=D=20$~km and $N(>D_p) = 4
\times 10^8$ (i.e. 4\% of the total number of particles of this size
in the disk) we find
$P^*=3.2\times$10$^{-8}$.  This means that the collisional lifetime is
$ 3.2\times$10$^{8}$ years, which is enormous relative to the
temporary capture time of $\sim 100$ y.  

In \citet{morbyetal2009}, the cumulative size frequency distribution
in the disk for $D<100$km is $(N > D_p) \propto D^{-2}$. In \citet{benzasphaug1999} $Q^*(D)$ $\propto D_t^{\beta}$ with
$\beta=1.25$. Moreover, from Eq. (3) one gets $D_p \propto (Q^*)^{1/3}
D_t$ and the cross-section of the target is $\propto
D_t^{2}$. Therefore, simple algebra shows that the collisional
lifetime of an object of size $D_T$ is proportional to
$D_T^{(2/3)\beta}$.  The total mass carried by the objects of this
size temporarily captured in the Pluto-Charon system decreases as
$1/D$. Thus, $D$ cannot be smaller than $D_{min}$=20km/160=0.125km for
which the collisional lifetime is 
$ 3.2\times10^{8}\times(0.125/20)^{0.83}\simeq5$My.
Thus, the negative conclusion achieved for 20km objects is valid at any size.

Therefore we are forced to conclude that the collisional break-up of
planetesimals temporary captured in the Pluto-Charon system cannot, by
orders of magnitude, deliver enough mass in debris to allow for the
subsequent formation of Nix and Hydra.

\section{Conclusions}

The origin of the small satellites of Pluto (Nix, Hydra and P4,5) is
still elusive. In particular, their distant and quasi-circular orbits
are difficult to explain in a scenario where these satellites are
envisioned to be small debris generated in the Charon-forming
collision \citep{lithwickwu2008,pealeetal2011,cheng2011}.

We have seen that planetesimals of the same size of
the current known satellites, or even larger, could
be captured in the Pluto-Charon system. However, the problem
with purely capture mechanism is that it generally produces 
satellites in high eccentric and inclined orbits, 
which eventually escape in a few 100 years. 
To trap these objects permanently and produce satellites on quasi-circular and coplanar orbits like Nix and Hydra,
we would need a strong damping mechanism.
Unfortunately, tidal interactions with Pluto could not circularize Nix and
Hydra's orbits even in a timescale as long as the age of
the solar system \citep{sternetal2006}.  
Thus, we miss a suitably strong damping mechanism.

Here, we have explored an alternative idea for the origin of these
satellites. Our conjecture was the following. When Pluto was still
embedded in a massive planetesimal disk, several planetesimals got
temporarily captured in the Pluto-Charon system. Some of
these planetesimals were subsequently disrupted by
collisions with other planetesimals on heliocentric orbits. The debris
generated by these disruptions formed a collisionally
dissipative disk, whose eccentricities and inclinations
eventually damped to zero. Thus the disk turned into an
accretional particle disk and the small satellites could be
formed.

We have explored quantitatively this idea. We found that temporary
capture of planetesimals in the Pluto-Charon system occurs with
non-negligible probability only if the planetesimal disk has a quite
low dynamical excitation (i.e. eccentricities only up to $\sim 0.03$
and inclinations up to half this value). However, the typical capture
time is only about 100y, much shorter than the collisional lifetime of
objects large enough to carry a sufficient amount of mass to form the
small Pluto satellites.  Thus, the captured objects should have
survived intact and should have not generated a disk of debris around
Pluto. Therefore, we conclude that the scenario that we envisioned is
not viable.

\citet{youdinetal2012}  have recently proposed an
alternative idea linked to the
possible formation of Pluto-Charon binary in a 
gravitational collapse scenario \citep{nesvornyetal2010}. 
In this framework, Pluto's outer
moons would have formed from a plutocentric disk 
composed of material leftover 
in the gravitational collapse process
on orbits bound to Pluto. The idea is appealing, but the accretion
of massive Kuiper belt binaries by collapsing swarms
of solids has yet to be investigated in details.

\begin{acknowledgements}
This work was done while the first author was on research stage at Observatoire de la C\^{o}te d'Azur, so P. M. Pires dos Santos thanks Fapesp for the financial
support.
\end{acknowledgements}

\bibliographystyle{spbasic}      % basic style, author-year citations
\bibliography{reference1}   % name your BibTeX data base

\end{document}